\documentclass[%
 reprint,
 superscriptaddress,
 showkeys,
nofootinbib,
 amsmath,amssymb,
 aps,
 prd,
]{revtex4-2}

\usepackage[utf8]{inputenc}
\DeclareUnicodeCharacter{2500}{\textemdash{}}

\usepackage{soul}
\usepackage{float}
\usepackage[caption=false]{subfig} 
\usepackage{gensymb}
\usepackage{graphicx}
\usepackage{dcolumn}
\usepackage{bm}
\usepackage[hypertexnames=true,colorlinks=true,urlcolor=blue,bookmarks=true,citecolor=blue,breaklinks=true]{hyperref}
\usepackage[dvipsnames]{xcolor}
\usepackage{booktabs}
\usepackage{makecell}
\usepackage{array}
\usepackage{multirow}
\usepackage[normalem]{ulem}
\newcolumntype{C}[1]{>{\centering\arraybackslash}p{#1}}

\begin{document}

\preprint{APS/123-QED}

\title{Parameter Estimation Horizon of Core-Collapse Supernovae with Current and Next-Generation Gravitational-Wave Detectors}

\author{A. Akhmetali}
\affiliation{Energetic Cosmos Laboratory, Nazarbayev University, 010000 Astana, Kazakhstan} 
\affiliation{Department of Electronics and Astrophysics, Al-Farabi Kazakh National University, 050040 Almaty,
Kazakhstan} 

\author{Y. S. Abylkairov}
\email{sultan.abylkairov@nu.edu.kz} 
\affiliation{Energetic Cosmos Laboratory, Nazarbayev University, 010000 Astana, Kazakhstan} 
\affiliation{School of Artificial Intelligence and Data Science, Astana IT University, 020000 Astana, Kazakhstan}

\author{D. Orel}
\affiliation{Department of Natural Language Processing, Mohamed bin Zayed University of Artificial Intelligence, Abu Dhabi 54115, United Arab
Emirates}

\author{S. Nunes}
\affiliation{Centro de Física das Universidades do Minho e do Porto (CF-UM-UP),
Universidade do Minho, 4710–057 Braga, Portugal}

\author{A. Sakan}
\affiliation{Department of Electronics and Astrophysics, Al-Farabi Kazakh National University, 050040 Almaty,
Kazakhstan} 
\affiliation{Energetic Cosmos Laboratory, Nazarbayev University, 010000 Astana, Kazakhstan}

\author{A. Zhunuskanov}
\affiliation{Department of Electronics and Astrophysics, Al-Farabi Kazakh National University, 050040 Almaty,
Kazakhstan} 

\author{M. Zaidyn}
\affiliation{Department of Electronics and Astrophysics, Al-Farabi Kazakh National University, 050040 Almaty,
Kazakhstan} 

\author{N. Ussipov}
\affiliation{Department of Electronics and Astrophysics, Al-Farabi Kazakh National University, 050040 Almaty,
Kazakhstan} 
\affiliation{Energetic Cosmos Laboratory, Nazarbayev University, 010000 Astana, Kazakhstan}

\author{J. A. Font}
\affiliation{Departamento de Astronomía y Astrofísica, Universitat de València, Avinguda Vicent Andrés Estellés 19, 46100 Burjassot (Valencia), Spain}
\affiliation{Observatori Astronòmic, Universitat de València, Catedrático José Beltrán 2, 46980, Paterna, Spain}

\author{E. Abdikamalov}
\affiliation{Energetic Cosmos Laboratory, Nazarbayev University, 010000 Astana, Kazakhstan}
\affiliation{Department of Physics, Nazarbayev University, 010000 Astana, Kazakhstan}

\date{\today}

\begin{abstract}
Core-collapse supernovae (CCSNe) are powerful sources of gravitational waves (GWs). These signals propagate essentially unobstructed, providing a unique probe of the supernova central engine. In this work, we investigate parameter estimation from the bounce and early ring-down GW signal of rotating CCSNe using machine learning. We infer the peak frequency and peak amplitude of the signal as well as the rotation of the core. We extend previous studies in several directions. We consider a range of progenitor models and nuclear equations of state, and we assess the impact of key physical uncertainties, including bounce time uncertainty and source inclination. We incorporate both current detector noise and the projected sensitivities of next-generation observatories. We find that uncertainty in the bounce time does not significantly affect parameter estimation when the analysis is performed in the Fourier domain. In contrast, orientations when the rotation axis is near the line of sight substantially degrade performance. For optimal orientations, next-generation detectors can constrain rotation out to distances exceeding 100 kpc.
\end{abstract}


\maketitle

\section{Introduction}
\label{sec:intro}

Gravitational waves (GWs) have become a transformative probe of the universe’s most extreme environments. Observations of compact binary mergers now provide insights into their formation and evolution, strong-field gravity, and the physics of dense matter \citep{Abac25GWTC4}. Core-collapse supernovae (CCSNe) represent a distinct class of GW emitters \citep{mezzacappa24gravitational, Vartanyan23Gravitational}. A direct detection would offer an unobstructed, real-time window into the supernova central engine, providing key insights into the physical mechanisms that drive the explosion \cite{Mueller26GWREview}. While current-generation detectors are well-positioned for a breakthrough discovery within the Milky Way \cite{abbott2020optically, szczepanczyk2023optically}, the enhanced sensitivity of future observatories will significantly enlarge the horizon distance of possible detections, increasing the event rate and providing much more insight into the physics of CCSNe~\citep{srivastava19detection}.

When a massive star exhausts its nuclear fuel, its core becomes unstable and collapses, initiating a CCSN. The collapse proceeds until nuclear densities are reached, at which point repulsive nuclear forces halt the infall, the core rebounds, and a shock wave is launched. However, the shock rapidly loses energy and stalls \cite{Janka16Physics}. For an explosion to occur, the shock must be re-energized and propagate outward, ultimately expelling the stellar envelope and leaving behind a compact neutron star \cite{Ertl16Twoparameter}. A fraction of the neutrinos emitted by the newly formed proto-neutron star (PNS) is absorbed in the region behind the shock, depositing energy that aids its revival \cite{Boccioli25Neutrino}. This process is known as the neutrino mechanism of CCSNe \cite{Janka01Conditions}. In rapidly rotating progenitors, rotational energy can also contribute to powering the explosion via the magnetorotational mechanism \cite{burrows:07b}. Despite decades of progress, the detailed mechanism of the explosion remains not fully understood \cite{burrows13colloquium}.

Observations of CCSNe across the electromagnetic spectrum are highly informative about the explosion outcome, but provide limited insight into the actual mechanism that initiates it. This is because the shock is energized within 1 s after core bounce, while still buried deep inside the stellar envelope and thus electromagnetically obscured \cite{oconnor11}. In contrast, neutrinos and GWs escape the core and directly probe the central engine. Neutrinos are emitted already at core bounce and over the first $\sim 10 $ s, while the GW signal arises within milliseconds to $\sim 1$ s after bounce, well before electromagnetic signals can emerge \cite{Nakamura:2016kkl, WesternacherSchneider19Multimessenger}. The focus of our investigation is on the information that can be inferred from the analysis of GWs emitted in the early-bounce signal. Since GWs are generated by accelerated mass motions, they encode information about the dynamics of the stellar matter \cite{mueller:13, Morozova18, TorresForne19Universal, powell22inferring}. Decades of numerical modeling have established a comprehensive understanding of the primary mechanisms driving GW emission in CCSNe. Rotation strongly influences the dynamics: slowly rotating or non-rotating models differ significantly from rapidly rotating ones. In slowly rotating models, neutrino heating and hydrodynamic instabilities driven by neutrino processes play a crucial role \cite{Burrows93, mueller:12, Radice16Neutrino}. These instabilities perturb the PNS and excite its oscillation modes, which produce the bulk of the GW emission \citep{Murphy09Model, Ott13General, Yakunin15GW, Andresen17Gravitational, TorresForne19Universal, Mezzacappa23Core, Vartanyan23Gravitational, Sotani24Universality, Ehring26Gravitational, Lella26Gravitational}. Pre-collapse convection in nuclear burning shells may further enhance this effect once these perturbations are advected into the post-shock region \cite{Couch13Revival, Couch15Three, Mueller17Supernova, nagakura:19, Kazeroni20impact, Vartanyan22collapse, Telman24Convective}. An additional contribution arises from non-radial shock oscillations, known as the standing accretion shock instability (SASI) \citep{blondin03stability, foglizzo06neutrino, mueller:12, kuroda16, kuroda:17, Buellet23Effect}. These hydrodynamic instabilities play a crucial role in driving the supernova explosion by enhancing the exposure to neutrino heating \citep[e.g.,][]{mueller:15nonradial, Bruenn16}.

In rapidly rotating models, which likely represent a minority of massive stars (possibly as low as $\sim 1 \%$) \citep{Heger05Presupernova, woosley:06}), the PNS is born centrifugally deformed. This deformation excites quadrupolar oscillations at core bounce that ring-down within $\sim 10$ ms \cite{ott12correlated, Fuller15SNseismology}. In some cases, the PNS can develop non-axisymmetric instabilities, leading to a non-axisymmetric shape that endures for several rotation periods and produces long-lasting GW emission \cite{Scheidegger08, Shibagaki20new, Pan21Stellar}. Additional contributions arise from anisotropic neutrino emission \citep{mueller:97, Takiwaki18Anisotropic, Vartanyan20GWanisotropy, Choi24GW}, asymmetric shock propagation \citep{radice:19gw}, and jet dynamics \citep{Birnholtz13GW_jet, Soker23GWJJ, Pais23choked}. In these models, the rotational energy of the PNS is transferred to the shock front—and thus to the explosion energy budget—via magnetic fields, in what is known as the magnetorotational mechanism \cite{moesta:14b, obergaulinger:20, kuroda:20}. We also note that GW emission from rotating progenitors may be amplified by a resonance between the PNS fundamental quadrupolar oscillation mode and epicyclic motions at the inner-core boundary~\citep{Cusinato:26} (see also \citep{Sykes25Trends}).

Once a GW signal is observed, a key goal is to infer the source parameters from the waveform \citep{pajkos21, abdikamalov22gravitational, CasallasLagos23Characterizing, mitra23}. Because rotation plays a major role in shaping the dynamics, the resulting GW emission carries clear signatures of the progenitor’s angular momentum. In particular, GW signals may distinguish between neutrino-driven explosions in slowly rotating progenitors and magnetorotational explosions in rapidly rotating ones \citep{Logue12Inferring, Powell24Determining}. They may also enable measurements of the core rotation \citep{abdikamalov:14, pajkos19} and place constraints on the PNS mass and radius \citep{Bizouard21Inference, Bruel23Inference}. In cases where the explosion fails, or when fallback accretion onto the PNS occurs, collapse to a black hole may follow \cite{cerda:13, Pan18Equation, Shibagaki21, Burrows23Black, Powell25noEMCCSN, Eggenberger25Black}, producing characteristic GW signatures \cite{Ott11PRL, Kuroda23Failed, Pan21Stellar}. Finally, GW observations may also constrain the EOS of high-density matter \cite{richers:17, edwards17, chao22determining, Wolfe23GW, mitra24, abylkairov2025assessing, Akhmetali26Toward, Murphy24Dependence, Rusakov26Exploration, Powell25Impact}.

In this work, we investigate parameter estimation for rapidly rotating CCSNe, focusing on the bounce and early post-bounce ring-down signals. This phase is highly amenable to numerical modeling, allowing for the generation of large waveform libraries at a modest computational cost \cite{ott:07cqg}, which enables machine learning analysis \cite{mitra23, mitra24, Abylkairov24Evaluating}. We infer three key parameters: the peak frequency $f_{\mathrm{peak}}$, the distance-normalized peak amplitude $D\Delta h$, and a rotation parameter at bounce. The first two characterize the GW signal: $D\Delta h$ is the difference between maximum and minimum strain (for optimal orientation), and $f_{\mathrm{peak}}$ is the highest frequency in the signal, reflecting PNS oscillations. Rotation is parametrized by $T/|W|$ at core bounce, the ratio of rotational kinetic to gravitational binding energy, which governs the dynamical impact of rotation and shapes the GW signal at bounce \cite{richers:17}.

These three parameters capture the core structure and bounce dynamics. The rotation parameter $T/|W|$ measures rotation and determines the resulting centrifugal flattening of the inner core. This deformation establishes the mass quadrupole moment, which affects the peak strain amplitude, $D\Delta h$. The GW amplitude reflects a competition between centrifugal deformation, which enhances the signal, and centrifugal slowing, which suppresses it. For $T/|W| \lesssim 0.05$, the former dominates, and the peak GW amplitude scales as $\propto T/|W|$ \cite{richers:17}. At higher $T/|W|$, centrifugal slowing becomes significant, leading to a saturation of $D\Delta h$ with increasing $T/|W|$. In this regime, the Coriolis force also becomes important and can excite inertial modes \cite{richers:17}. As a result, both the GW amplitude and the frequency exhibit a nontrivial dependence on $T/|W|$. Therefore, measuring rotation parameter $T/|W|$ also probes the dynamics of the PNS. Moreover, while $D\Delta h$ is the peak amplitude, $f_{\mathrm{peak}}$ captures the dominant feature of the spectrum, so measuring both gives a more complete characterization of the GW signal. The frequency $f_{\mathrm{peak}}$ reflects the oscillation frequency of the PNS ring-down. Because $f_{\mathrm{peak}}$ depends on the PNS structure \citep{richers:17}, measuring $f_{\mathrm{peak}}$ alongside $D\Delta h$ helps to break the degeneracy between $D\Delta h$ and $T/|W|$. This motivates independently inferring $D\Delta h$, $f_{\mathrm{peak}}$, and $T/|W|$ from the signal.

We build on the recent works by \citep{Edwards2014inference, chao22determining, pastor24, nunes2024deep, Villegas25Parameter}. We extend those studies by incorporating a broader set of physical and observational variables. Specifically, we include four progenitor models to capture variations in stellar structure, and a diverse set of nuclear equations of state (EOS) to quantify their impact on parameter inference. Moreover, we account for uncertainty in the bounce time—difficult to constrain from GW and neutrino signals alone \citep{Pagliaroli09, Halzen09}—and explore the effect of source inclination on our results. We assess the performance of the inferences using both current GW detectors, namely Advanced LIGO (A+)~\cite{Hild2011ET}, and the next-generation detectors Einstein Telescope (ET)~\cite{2026:ET} and Cosmic Explorer (CE)~\cite{hall_cosmic_2022}. For the latter, we adopt their projected sensitivities. Finally, we complement time-domain (TD) inference with frequency-domain (FD) analyses to exploit the strengths of both data representations.

The following sections outline the remainder of this work: Section~\ref{sec:methods} details the dataset and methodological framework employed. In Section~\ref{sec:results}, we present our primary findings and analysis. Finally, Section~\ref{sec:conclusion} provides a summary of our core results and concluding remarks.

\section{Methods}
\label{sec:methods}

Below we describe the dataset, the preprocessing steps of the data, and the regression models used to estimate the physical parameters of the source from its GW signals.

\subsection{Data}
\label{sec:data}

The catalog used here is identical to that in our previous work \citep{Akhmetali26Toward}\footnote{The data are available at https://zenodo.org/records/17579189}; we refer the reader to \citep{Akhmetali26Toward} for full details and summarize only the key elements below. GW signals are computed with the relativistic {\tt CoCoNuT} code \citep{Dimmelmeier02a, dimmelmeier:05MdM}. Simulations are axisymmetric, which is adequate through collapse, bounce, and early post-bounce \citep{ott:07cqg}. We use the electron fraction parametrized as a function of density, $Y_e(\rho)$, to produce the deleptonization scheme during collapse \citep{Liebendoerfer05Simple} and a leakage/heating scheme after bounce \citep{Ott13General}. We consider four solar-metallicity progenitor models with ZAMS masses of $12$, $15$, $27$, and $40\,M_\odot$ \citep{woosley:07, Heger05Presupernova, whw:02, woosley:07}, spanning a range of core structures. Moreover, we use six EOSs (\texttt{BHB}$\Lambda \Phi$ \cite{bhbeos}, \texttt{SFHo} \cite{steiner:13b}, \texttt{SFHx} \cite{steiner:13b}, \texttt{LS220} \cite{lseos:91}, \texttt{HSDD2} \cite{hempel:10,hempel:12}, \texttt{GShenFSU2.1} \cite{gshen:11b}). For each mass–EOS pair, we sample rotation over the range $T/|W| \in [0.02, 0.21]$, spanning slow to rapid rotation, yielding 1332 waveforms in total (4096 Hz sampling rate). The analysis is restricted to a $[-2,6]$ ms window around bounce to isolate the bounce and early ring-down signal; later times are avoided due to contribution of the prompt convection, which is poorly captured in 2D. Our tests show that inclusion of the prompt convection has minimal impact on the parameter estimation accuracy, suggesting the model is driven primarily by the bounce/ring-down features.

\subsection{Signal injection and data preprocessing}
\label{sec:injection}

For the detector noise, we use publicly available data from the A+ Hanford detector obtained from the latest O4a public release~\citep{O4a}, spanning GPS times $t_{\rm GPS}=$ 1378195220--1378196244 s\footnote{We note that detector noise properties may vary on secular timescales due to changes in instrumental and environmental conditions.}, corresponding to early September 2023. The data are sampled at 4096~Hz, and a glitch-free 1024~s segment of detector noise is used for signal injection. In addition to the current-generation detector, we also consider the sensitivity of third-generation GW observatories. For this purpose, simulated Gaussian noise is generated using the design sensitivity curves of  ET\footnote{We use \texttt{EinsteinTelescopeP1600143} PSD from \texttt{PyCBC}} and CE\footnote{We use \texttt{CosmicExplorerP1600143} PSD from \texttt{PyCBC}} with the \texttt{PyCBC} library~\citep{pycbc}, allowing us to assess the performance of the proposed method for future detectors. We adopt idealized conditions, assuming that the detector arms are optimally oriented relative to the signal.

We inject the signals into detector noise in the TD. Prior to injection, each waveform is smoothly tapered using a Tukey window~\citep{tukey1967intoroduction} ($\alpha=0.1$) to suppress edge effects and reduce spectral leakage. After injection, the resulting strain data are whitened using the power spectral density (PSD) of the detector noise. The strain is then bandpass filtered between 20 and 2000~Hz. For the A+ data, an additional notch filter is applied at the power-line harmonics of 60, 120, and 240~Hz to suppress instrumental contamination. From the processed strain, we construct two types of input representations for the machine learning models: the TD strain and its FD representation obtained via Fourier transform. 

For the regression analysis, the resulting strain is segmented into 30~ms windows. As noted before, the injected GW signals have a duration of 8~ms. To account for uncertainty in the core-bounce time, the signal is injected at a random position within the window, between the beginning of the segment and a time offset $\Delta t_{\rm b}$. In this work, we consider two cases of bounce time uncertainty, $\Delta t_{\rm b}=0$~ms and $\Delta t_{\rm b}=20$~ms. For all configurations, the catalog is split into training and testing datasets, with 80\% used for training and the remaining 20\% used for evaluation. To ensure that the reported results are not dependent on a particular train–test split, the evaluation procedure was repeated 50 times. In each iteration, a new random train–test split was generated. The final reported values correspond to the mean and standard deviation computed across all 50 iterations. This procedure was used for the results presented in Sections \ref{sec:bouncetime}, \ref{sec:eos}, \ref{sec:mass}, and \ref{sec:inclination}.

\subsection{Machine learning}
\label{sec:MLmodels}

In regression tasks, selecting an appropriate machine learning model is an important step, as different algorithms exhibit different inductive biases and predictive capabilities~\citep{alzubi2018machine}. In this work, we evaluate the performance of seven classical regression models for parameter estimation and determine which approach is most suitable for our task. Specifically, we employ Ridge regression~\citep{hoerl1970ridge}, Lasso regression~\citep{tibshirani1996regression}, Random Forest Regressor (RFR)~\citep{breiman2001random}, Support Vector Regression (SVR)~\citep{smola2004tutorial}, CatBoost~\citep{catboost}, LightGBM (LGB)~\citep{lgb}, and the XGBoost Regressor (XGB)~\citep{chen2016xgboost}. We use {\tt scikit-learn}~\citep{pedregosa2011scikit}, {\tt catboost}~\citep{catboost}, {\tt lightgbm}~\citep{lightgbm_manual}, and {\tt xgboost}~\citep{xgboost_manual} libraries to implement the models.

In all regression experiments, each physical parameter is modeled independently. That is, separate machine learning models are trained and evaluated for each target parameter, rather than performing a joint multi-output regression.

Before evaluating their predictive performance, each model is optimized with respect to its hyperparameters in order to achieve its best possible performance on the dataset. This is done via grid search with cross-validation~\citep{pedregosa2011scikit}. This method performs an exhaustive search over a predefined grid of hyperparameter values and evaluates each configuration using cross-validation. In our case, a 5-fold cross-validation scheme is used, where the training dataset is divided into 5 subsets. For each hyperparameter combination, the model is trained on four subsets and validated on the remaining subset. This procedure is repeated five times so that each subset is used once as a validation set. The performance scores from all folds are then averaged to provide a robust estimate of the model's generalization capability. The hyperparameter combination that yields the best average score is selected as the optimal configuration. The resulting optimal hyperparameters are then fixed for all subsequent analyses. Model performance is subsequently evaluated using 50 independent resampling runs, which are not involved in the hyperparameter optimization process.

The hyperparameter grids used for model tuning are summarized in Table~\ref{tab:Table 1}, with the optimal values highlighted in bold. The models with these tuned hyperparameters are then used in the subsequent regression analysis to assess their ability to recover the physical parameters from the gravitational wave signals. 

We note that the selected grids focus on the most influential hyperparameters for each model, while remaining parameters are kept at their default values to avoid an excessively large search space. Although some optimal values lie near the boundaries of the explored grids, this work is not aimed at exhaustive hyperparameter optimization; rather, the goal is a consistent and fair comparison across models under a controlled tuning strategy. A finer or expanded search could potentially yield marginal improvements, but is not expected to qualitatively change the comparative trends observed in this study.

\begin{table}[h]
\centering
\caption{Hyperparameters grid used for tuning machine learning models. The optimal hyperparameters are highlighted in bold.}
\label{tab:Table 1}
\begin{tabular}{ll}
\toprule
Model & Hyperparameter grid \\ \midrule
Ridge         & alpha: [0.1, 1.0, \textbf{10.0}, 100.0] \\ \midrule
Lasso         & alpha: [\textbf{0.0001}, 0.001, 0.01, 0.1, 1.0, 10.0] \\ \midrule
RFR           & n\_estimators: [\textbf{100}, 300] \\
              & max\_depth: [10, \textbf{20}, None] \\
              & min\_samples\_split: [\textbf{2}, 5, 10] \\
              & max\_features: [\textbf{sqrt}, log2] \\ \midrule
SVR           & C: [0.1, \textbf{1}, 10, 100] \\
              & gamma: [\textbf{scale}, auto, 0.01, 0.1] \\
              & epsilon: [\textbf{0.01}, 0.1, 0.2] \\ \midrule
CatBoost      & iterations: [100, 300, \textbf{500}] \\
              & depth: [2, \textbf{4}, 6] \\
              & learning\_rate: [0.01, \textbf{0.1}, 0.2] \\ \midrule
LGB           & n\_estimators: [100, \textbf{300}, 500] \\
              & learning\_rate: [0.01, \textbf{0.1}, 0.2] \\
              & num\_leaves: [31, \textbf{63}, 127] \\
              & subsample: [\textbf{0.8}, 1.0] \\ \midrule
XGB           & n\_estimators: [100, 300, \textbf{500}] \\
              & max\_depth: [\textbf{3}, 6, 10] \\
              & learning\_rate: [0.01, \textbf{0.1}, 0.2] \\
              & subsample: [\textbf{0.8}, 1.0] \\ \bottomrule
\end{tabular}
\end{table}

We evaluate the parameter estimation performance of our models using the coefficient of determination (\(R^2\)) score, defined as:
\[
R^2 = 1 - \frac{\sum_{i=1}^{N}(y_i-\hat{y}_i)^2}
{\sum_{i=1}^{N}(y_i-\bar{y})^2},
\]
where \(y_i\) and \(\hat{y}_i\) denote the true and predicted values of the target parameter for the \(i\)-th sample, respectively, \(\bar{y}\) is the mean of the true values, and \(N\) is the total number of samples. The \(R^2\) metric quantifies the proportion of the variance in the target parameter that is explained by the model predictions, making it a widely used measure of regression performance. An \(R^2\) value of \(1\) indicates perfect agreement between predictions and observations, whereas an \(R^2\) value of \(0\) implies that the model performs no better than predicting the mean of the target variable. Negative values indicate that the model performs worse than this baseline. In general, \(R^2\) values above \(0.9\) are considered excellent, while values between \(0.8\) and \(0.9\) indicate strong predictive capability~\cite{coulibaly_r2}.

\section{Results}
\label{sec:results}

\subsection{Machine learning models}
\label{sec:ML}

We first evaluate several machine learning models to compare their performance in estimating the physical parameters of gravitational waves on their TD representation. We use the hyperparameter-tuned models described in Section~\ref{sec:MLmodels} and test them with GW signals injected at a reference distance of 10~kpc, corresponding approximately to the Galactic center. We assume an optimally oriented source at a fixed optimal sky position, and adopt the A+ detector noise configuration for this test.

Table~\ref{tab:Table 2} summarizes the $R^2$ scores obtained for each machine learning model across the three estimated physical parameters. Overall, the gradient boosting methods outperform the other approaches, with XGB and LGB consistently achieving the highest scores. In particular, XGB yields the best performance for $f_{\mathrm{peak}}$ with $R^2$ value of 0.85, while sharing the best performance for $T/|W|$ and $D\Delta h$ with $R^2$ values of 0.99 and 0.97, respectively. CatBoost also performs competitively, with $R^2$ values of 0.98, 0.85, and 0.97 for $T/|W|$, $f_{\mathrm{peak}}$, and $D\Delta h$, respectively.

In contrast, the linear models (Ridge and Lasso) yield consistently lower $R^2$ scores than the boosting methods, although they still achieve strong predictive performance, with $R^2$ values of 0.97 for $T/|W|$, 0.80--0.81 for $f_{\mathrm{peak}}$, and 0.97 for $D\Delta h$. Meanwhile, RFR and SVR exhibit comparatively weaker performance, particularly for $f_{\mathrm{peak}}$, where the $R^2$ scores decrease to 0.76 and 0.81, respectively. For $T/|W|$, both methods achieve $R^2$ values of 0.96, while for $D\Delta h$ they attain $R^2$ values of 0.91 and 0.95, respectively. These trends highlight the advantage of boosting-based models for capturing the underlying relationships in the data. Based on these results, we select XGB as the primary model for the subsequent regression analysis. We note that similar trends are observed for the FD representation and for non-zero values of the $\Delta t_b$.

\begin{table}[htbp]
\centering
\caption{$R^2$ scores of seven classical machine learning models for estimating physical parameters from TD signals at a reference distance of 10~kpc (Galactic center) with $\Delta t_{\mathrm b}=0$~ms. For each parameter (row), bold values indicate the highest $R^2$ across the seven models.}
\label{tab:Table 2}
\begin{tabular}{@{}lccccccc@{}}
\toprule
Parameter & Ridge & Lasso & RFR & SVR & CatBoost & LGB & XGB \\ \midrule
$T/|W|$              & 0.97 & 0.97 & 0.96 & 0.96 & 0.98 & {\bf 0.99} & {\bf 0.99} \\
$f_{\mathrm{peak}}$  & 0.81 & 0.80 & 0.76  & 0.81  & 0.82 & 0.84 & {\bf 0.85} \\
$D\Delta h$          & {\bf 0.97} & {\bf 0.97} & 0.91 & 0.95  & {\bf 0.97} & {\bf 0.97} & {\bf 0.97} \\ \bottomrule
\end{tabular}
\end{table}

\subsection{Impact of Bounce Time Uncertainty} 
\label{sec:bouncetime}

In a realistic observing scenario, the exact moment of the core-collapse bounce is difficult to determine. The GW signal alone does not provide sufficiently precise timing information. Moreover, even with a high-statistics neutrino detection, the bounce time can only be constrained to within a window of approximately $10\,\mathrm{ms}$~\citep{Pagliaroli09, Halzen09}. To account for these practical limitations, we introduce a bounce time uncertainty of 20\,ms and evaluate the performance of our models in both TD and FD data representations.

The results, again at a reference distance of 10\,kpc and injected in real A+ detector noise, are summarized in Table~\ref{tab:Table 3}. When the bounce time is known exactly ($\Delta t_{\mathrm{b}} = 0$\,ms), both feature representations achieve strong performance, with $R^2$ scores above $0.83$ for all parameters. However, introducing a 20\,ms timing uncertainty significantly degrades the performance of the TD-based model. In particular, the  $R^2$ scores for $T/|W|$, $f_{\mathrm{peak}}$, and $D\Delta h$ decrease from $0.99$, $0.83$, and $0.96$ to $0.74$, $0.30$ and $0.53$, respectively. This decline indicates that TD features are highly sensitive to the precise alignment of the waveform within the analysis window. We note that a similar trend was observed for the TD-based model in the context of EOS classification in~\citep{Akhmetali26Toward}, where performance also deteriorates with increasing bounce time uncertainty.

In contrast, the FD-based model shows almost no degradation in performance under the same timing uncertainty. The $R^2$ scores for all parameters remain nearly unchanged. This robustness arises from the properties of the Fourier transform: a time shift alters the phase of the signal but does not affect the magnitude of its frequency spectrum. Since our FD features are derived from the spectral magnitude, the model can still extract the relevant physical information even when the signal timing is uncertain. These results indicate that FD features provide a more reliable representation for CCSNe GW data analysis pipelines, as they remove the need for sub-millisecond timing precision. Therefore, in the following analysis we adopt the FD-based model with a bounce time uncertainty of $\Delta t_{\mathrm b}=20$\,ms. We note that this comes at the cost of discarding phase information, which may encode additional temporal structure of the signal; however, in this work we focus on amplitude-based spectral features for robustness to timing uncertainties.

\begin{table}[t]
\centering
\caption{$R^2$ scores of different input representations under bounce time uncertainty for a source at 10\,kpc. TD denotes the time-domain representation, while FD denotes the frequency-domain representation. Metrics represent the mean $R^2$ score ($\mu$) and standard deviation ($\sigma$) over 50 runs.}
\label{tab:Table 3}
\resizebox{\columnwidth}{!}{
\begin{tabular}{@{}lcccc@{}}
\toprule
& \multicolumn{2}{c}{$\Delta t_{\mathrm b}=0$\,ms} & \multicolumn{2}{c}{$\Delta t_{\mathrm b}=20$\,ms} \\
\cmidrule(lr){2-3} \cmidrule(l){4-5}
Parameter & TD & FD & TD & FD \\ \midrule
$T/|W|$         & $0.99 \pm 0.00$ & $0.98 \pm 0.00$ & $0.74 \pm 0.02$ & $0.98 \pm 0.00$ \\
$f_{\mathrm{peak}}$ & $0.83 \pm 0.03$ & $0.84 \pm 0.03$ & $0.30 \pm 0.06$ & $0.87 \pm 0.02$ \\
$D\Delta h$     & $0.96 \pm 0.00$ & $0.97 \pm 0.00$ & $0.53 \pm 0.04$ & $0.97 \pm 0.00$ \\ \bottomrule
\end{tabular}
}
\end{table}

\begin{figure*}[ht]
\centering
\includegraphics[width=1\linewidth]{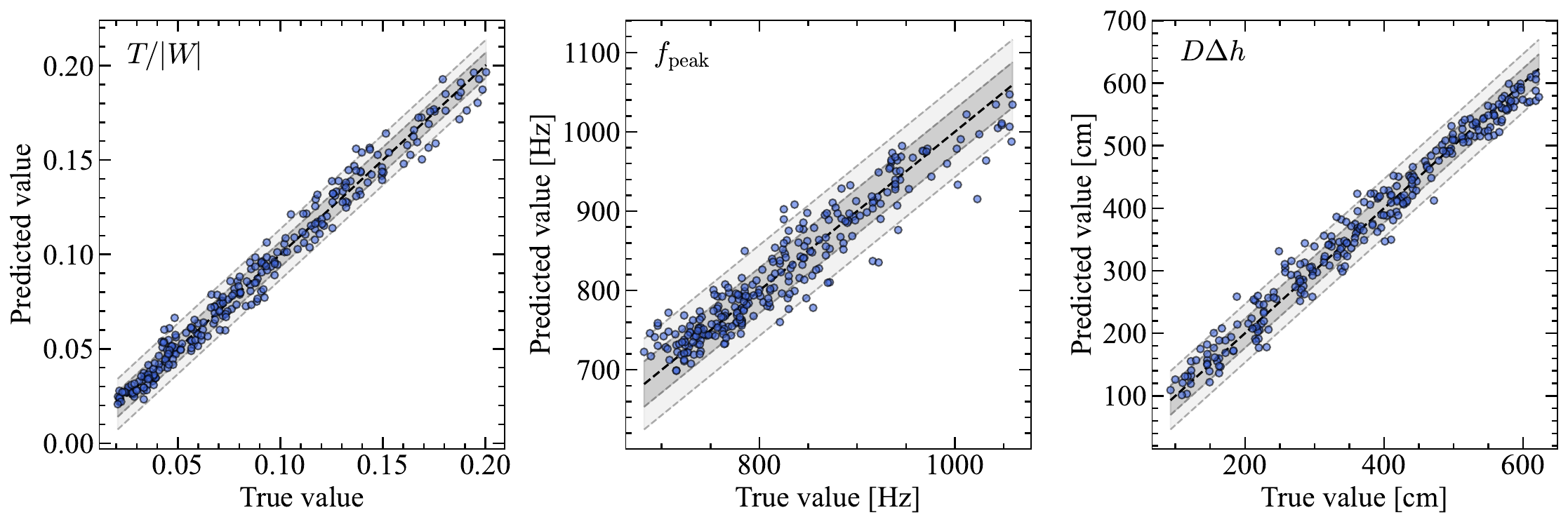}
    \caption{True vs. predicted values of the estimated parameters using the FD representation with $\Delta t_{\mathrm b}=20$\,ms and signals at a distance of 10\,kpc. The black dashed diagonal line corresponds to the ideal prediction. Gray dashed lines and shaded regions indicate the $1\sigma$ and $2\sigma$ deviation intervals. $R^2$ scores are $0.98$, $0.90$, and $0.97$} for $T/|W|$, $f_\mathrm{peak}$, and $D\Delta h$, respectively.
     \label{fig:true-pred}
\end{figure*}

Figure~\ref{fig:true-pred} shows the relationship between the true and predicted values of the estimated parameters using the FD-based model with \(\Delta t_{\mathrm b}=20\)\,ms for signals at a distance of 10\,kpc, assuming an optimally oriented source at a fixed optimal sky position and A+ detector noise. The predictions closely follow the diagonal line corresponding to perfect agreement, with $R^2$ scores of $0.98$, $0.90$, and $0.97$ for $T/|W|$, $f_{\mathrm{peak}}$ and $D \Delta h$, respectively. The majority of the predictions lie within the \(2\sigma\) confidence region, demonstrating that the model not only achieves accurate point estimates but also maintains a reliable level of statistical consistency.

\begin{figure}[h!]
\centering
\includegraphics[width=0.9\linewidth]{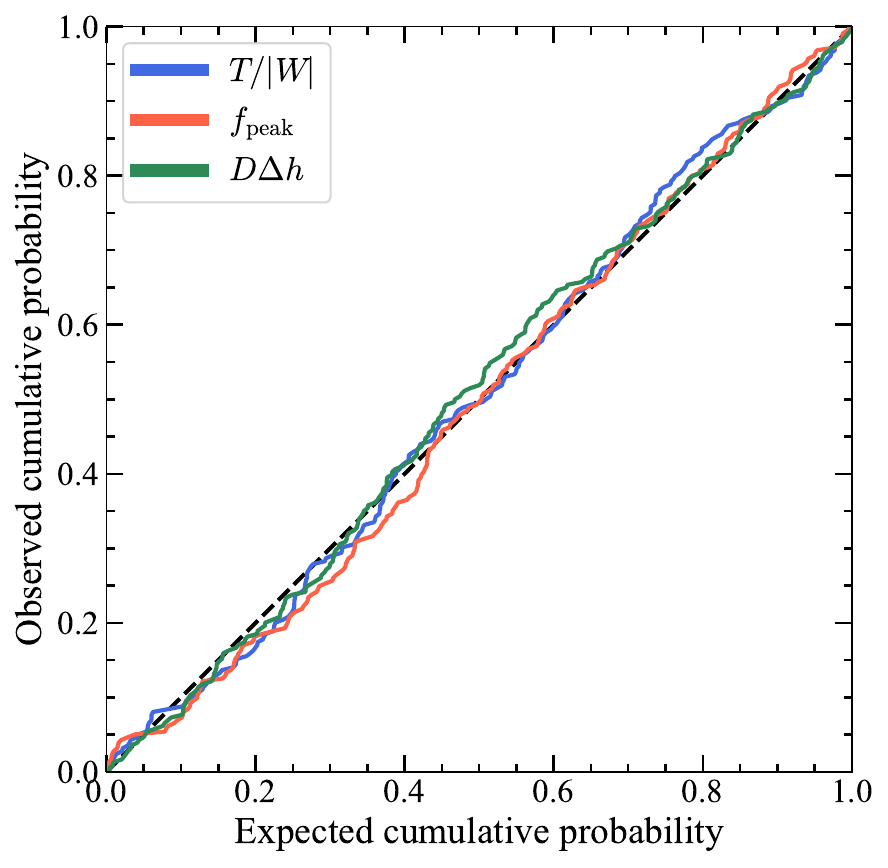}
    \caption{P-P plot for the estimated parameters at 10\,kpc using the FD-based model. The curve shows the fraction of true parameter values contained within the predicted credible intervals as a function of the nominal confidence level. The dashed diagonal line indicates perfect calibration.} 
     \label{fig:p-p}
\end{figure}

To further evaluate the reliability of the parameter estimation, we examine the probability–probability (P–P) plot, which is shown in Fig.~\ref{fig:p-p}. The P–P plot compares the cumulative distribution of the predicted probabilities with the expected uniform distribution. If the uncertainties are well calibrated, the curve should follow the diagonal line corresponding to perfect agreement. As seen in Fig.~\ref{fig:p-p}, the curves closely track the diagonal, indicating good calibration of the predicted uncertainties. This is quantified using the Kolmogorov--Smirnov (KS) statistic~\citep{KStest}, which measures the maximum deviation from the ideal diagonal. We obtain KS values of $0.039$ for $T/|W|$, $0.047$ for $f_{\mathrm{peak}}$, and $0.043$ for $D\Delta h$. These small deviations ($\lesssim 0.05$) indicate that the predicted distributions are statistically consistent with the expected calibration and that the model provides reliable probabilistic estimates of the parameters. 

\subsection{Impact of EOS}
\label{sec:eos}

We now analyze how the nuclear EOS affects the parameter estimation performance. Different EOSs leads to slightly different GW signals (see~\cite{richers:17} for details) and this may affect the accuracy of the inferences if the model is trained on one EOS and tested on another. To this end, we test whether the model can generalize to signals generated with an EOS that was not included during training. For this purpose, we adopt a hold-one-out strategy, in which the model is trained on data from $N-1$ EOSs, where $N$ is the total number of EOSs, and evaluated on the remaining unseen EOS. 

Table~\ref{tab:Table 4} summarizes the regression performance for each held-out EOS. The model demonstrates strong generalization, with $R^2$ scores remaining consistently high across EOSs: $T/|W|$ ranges from $0.96$ to $0.98$, $f_{\mathrm{peak}}$ from $0.84$ to $0.87$, and $D\Delta h$ from $0.97$ to $0.98$. These results indicate that the predictions are stable and largely independent of the EOS used for training. This behavior is expected, as the EOS has only a modest impact on the overall signal (see~\cite{richers:17}) and does not significantly affect the dominant features from which $T/|W|$, $f_{\mathrm{peak}}$, and $D\Delta h$ are inferred.

\begin{table}[htbp]
\centering
\caption{Model generalization performance across different held-out EOSs. Metrics represent the mean $R^2$ score ($\mu$) and standard deviation ($\sigma$) over 50 runs.}
\label{tab:Table 4}
\begin{tabular}{@{}lccc@{}}
\toprule
Held-out EOS & $T/|W|$ & $f_{\mathrm{peak}}$ & $D\Delta h$ \\ \midrule
\texttt{BHB} & $0.98 \pm 0.00$ & $0.86 \pm 0.01$ & $0.97 \pm 0.00$ \\
\texttt{SFHx} & $0.98 \pm 0.00$ & $0.86 \pm 0.01$ & $0.97 \pm 0.00$ \\
\texttt{SFHo} & $0.96 \pm 0.01$ & $0.88 \pm 0.01$ & $0.97 \pm 0.00$ \\
\texttt{GShenFSU2.1} & $0.98 \pm 0.00$ & $0.87 \pm 0.01$ & $0.97 \pm 0.00$ \\
\texttt{HSDD2}  & $0.98 \pm 0.00$ & $0.87 \pm 0.01$ & $0.97 \pm 0.00$ \\
\texttt{LS220} & $0.98 \pm 0.00$ & $0.84 \pm 0.02$ & $0.98 \pm 0.00$ \\ \bottomrule
\end{tabular}
\end{table}

\subsection{Impact of Progenitor Mass}
\label{sec:mass}

We now analyze how the progenitor mass affects the parameter estimation. If the model is trained on one progenitor mass and tested on another one, this may affect the performance of the parameters estimation. Similar to the EOS study above, we adopt a hold-one-out strategy, training the model on all but one progenitor mass and evaluating performance on the held-out mass.

Table~\ref{tab:Table 5} summarizes the regression performance for each held-out progenitor mass. As for the case of the EOS, the model demonstrates strong generalization, with $R^2$ scores remaining consistently high across all masses: $T/|W|$ stays at $0.98$, $f_{\mathrm{peak}}$ ranges from $0.85$ to $0.87$, and $D\Delta h$ from $0.97$ to $0.98$. These results indicate that the predictions are stable and largely independent of the progenitor mass used for training. This is again not surprising as different progenitors produce similar bounce signal for the same specific angular momentum distribution with the same enclosed mass \cite{mueller:09phd, ott12correlated, mitra23, Sakan25Probing}. 

\begin{table}[htbp]
\centering
\caption{Model generalization performance across different held-out progenitor masses. Metrics represent the mean $R^2$ score ($\mu$) and standard deviation ($\sigma$) over 50 runs.}
\label{tab:Table 5}
\begin{tabular}{@{}lccc@{}}
\toprule
Held-out Mass & $T/|W|$ & $f_{\mathrm{peak}}$ & $D\Delta h$ \\ \midrule
$12\,M_{\odot}$                      & $0.98 \pm 0.00$ & $0.85 \pm 0.01$ & $0.97 \pm 0.00$ \\
$15\,M_{\odot}$  & $0.98 \pm 0.00$ & $0.86 \pm 0.02$ & $0.97 \pm 0.00$ \\
$27\,M_{\odot}$  & $0.98 \pm 0.00$ & $0.86 \pm 0.01$  & $0.98 \pm 0.00$ \\
$40\,M_{\odot}$  & $0.98 \pm 0.00$ & $0.87 \pm 0.01$ & $0.97 \pm 0.00$ \\ \bottomrule
\end{tabular}
\end{table}

\subsection{Impact of Inclination Angle}
\label{sec:inclination}

We now examine how the inclination angle of the source affects parameter estimation. The dynamics during core bounce and the early postbounce phase are largely axisymmetric (see e.g.~\cite{ott12correlated}). For axisymmetric GW sources, the gravitational-wave strain scales as $\propto \sin^2\theta$, where $\theta$ is the inclination angle between the rotation axis and the line of sight. As a result, the observed signal amplitude decreases with the inclination angle, which can impact the regression performance of the model.

In this experiment, we train the model on a dataset of randomly oriented sources and evaluate it on signals with a fixed inclination angle $\theta$. The regression performance for different inclination angles is summarized in Table~\ref{tab:Table 6}. Overall, the results show a clear dependence of the predictive performance on the inclination angle, with the lowest accuracy observed at small inclinations ($\theta \leq 20^\circ$). In this regime, the performance degrades significantly, particularly for $f_{\mathrm{peak}}$ and $D\Delta h$, where the $R^2$ scores drop to $0.53 \pm 0.09$ and $0.66 \pm 0.09$ at $\theta = 10^\circ$, respectively, indicating reduced predictive reliability. The estimation of $T/|W|$ is also strongly affected at very low inclinations, reaching $R^2 = 0.79 \pm 0.03$ at $\theta = 10^\circ$, although it improves rapidly with increasing angle.

% We observe that the dependence of parameter estimation performance on inclination is not strictly monotonic. At very low inclination angles, the waveform becomes more axisymmetric, reducing the diversity of observable features and introducing partial degeneracies between different source parameters. As the inclination increases, the predictive performance improves across all targets. At $\theta = 30^\circ$, all quantities already reach strong accuracy, with $R^2$ values of $0.93 \pm 0.01$ for $T/|W|$, $0.85 \pm 0.03$ for $f_{\mathrm{peak}}$, and $0.83 \pm 0.01$ for $D\Delta h$. For moderate to large inclinations ($\theta \geq 45^\circ$), the model achieves consistently excellent performance, with $R^2$ exceeding $0.90$ for all parameters.

These results demonstrate that the model provides reliable and largely orientation-independent parameter estimates for moderate to high inclination angles, where the observed strain amplitude is sufficiently strong. In contrast, at low inclinations, the gravitational-wave signal is significantly suppressed due to projection effects, reducing the available information content and leading to degraded parameter recovery, particularly for amplitude-dependent quantities such as $f_{\mathrm{peak}}$ and $D\Delta h$.

\begin{table}[htbp]
\centering
\caption{Model performance across various source inclination angles. Metrics represent the mean $R^2$ score ($\mu$) and standard deviation ($\sigma$) over 50 runs.}
\label{tab:Table 6}
\begin{tabular}{@{}lccc@{}}
\toprule
$\theta$ & $T/|W|$ & $f_{\mathrm{peak}}$ & $D\Delta h$ \\ \midrule
$10^{\circ}$   & $0.79 \pm 0.03$ & $0.53 \pm 0.09$ & $0.66 \pm 0.09$ \\
$20^{\circ}$   & $0.87 \pm 0.02$ & $0.81 \pm 0.02$ & $0.74 \pm 0.03$ \\
$30^{\circ}$   & $0.93 \pm 0.01$  & $0.85 \pm 0.03$ & $0.83 \pm 0.01$  \\
$45^{\circ}$   & $0.96 \pm 0.00$ & $0.91 \pm 0.02$  & $0.90 \pm 0.01$ \\
$60^{\circ}$   & $0.98 \pm 0.00$ & $0.92 \pm 0.02$  & $0.94 \pm 0.01$ \\
$75^{\circ}$   & $0.98 \pm 0.00$ & $0.94 \pm 0.02$ & $0.96 \pm 0.01$  \\
$90^{\circ}$   & $0.97 \pm 0.00$  & $0.93 \pm 0.03$ & $0.94 \pm 0.01$ \\ \bottomrule
\end{tabular}
\end{table}

\subsection{Parameter estimation distance}
\label{sec:distance}

Finally, we investigate the maximum distances at which our set of physical parameters can be reliably estimated. As the distance increases, the observed GW strain decreases, reducing the signal-to-noise ratio and making parameter recovery more challenging. Consequently, the regression performance is strongly determined by the sensitivity of the detector.

In this experiment, we train and evaluate the model separately at each distance in the range 0.1--2000\,kpc. For each fixed distance, we construct a dedicated dataset, and an independent model is trained and tested without mixing samples from other distances. This setup allows us to isolate the dependence of performance on distance-driven SNR scaling. The distance values are sampled on a logarithmically spaced grid, providing higher resolution at small distances where the parameter estimation performance changes rapidly, while more sparsely sampling the large-distance regime where the variation is smoother.

\begin{figure*}[htbp]
\centering
\includegraphics[width=1\linewidth]{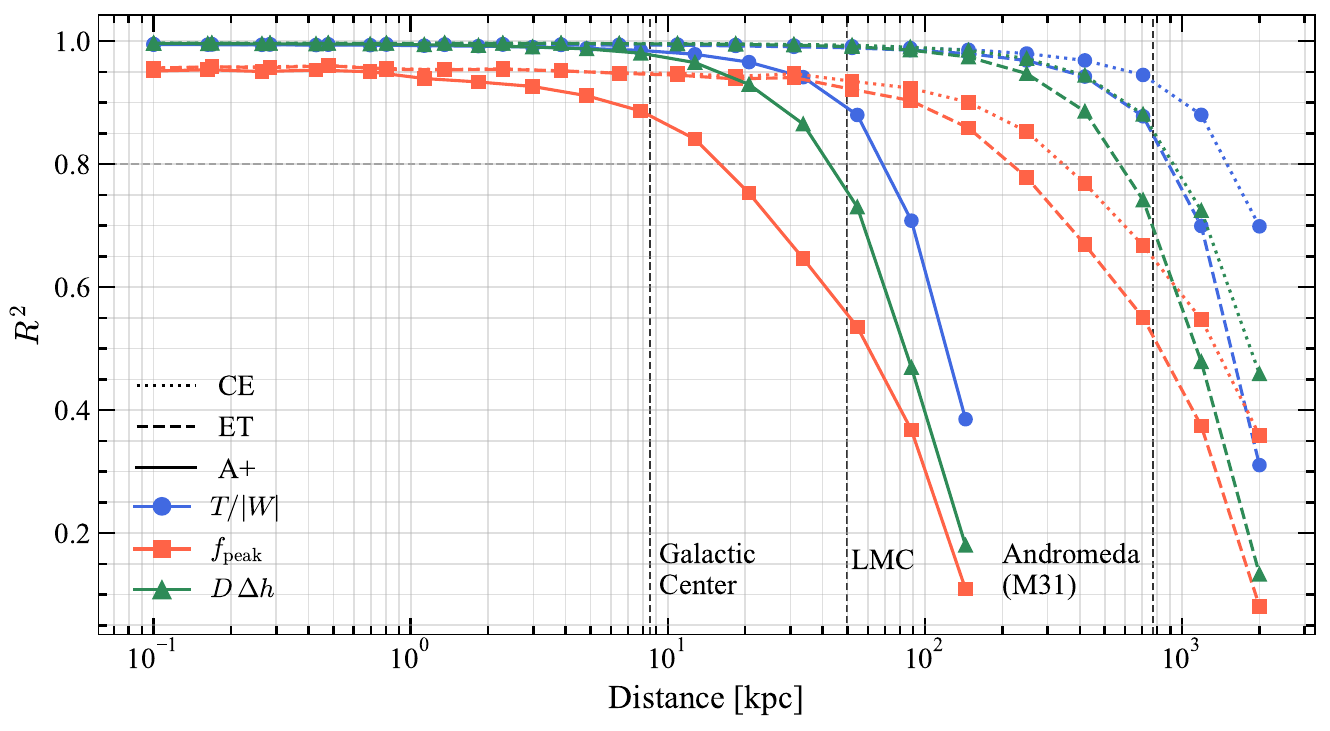}
    \caption{$R^2$ score as a function of source distance for different parameters and detectors. The blue, orange, and green lines correspond to $T/|W|$, $f_{\mathrm{peak}}$, and $D\Delta h$, respectively. Solid lines represent the A+ detector, while dashed and dotted lines correspond to the ET and CE 3G detectors. The gray horizontal dashed line at 0.8 indicates a threshold for reasonably accurate practical predictions. Vertical black dashed lines mark reference distances to the Galactic Center (8.5\,kpc), the Large Magellanic Cloud (49.8\,kpc), and M31 (780\,kpc).}
     \label{fig:fraction}
\end{figure*}

Figure~\ref{fig:fraction} shows $R^2$ score as a function of distance for both second-generation (2G) and third-generation (3G) detectors. For the A+ detector, high-precision parameter estimation is achievable only for nearby sources, with reliable recovery limited to distances of approximately $\sim 15$\,kpc, corresponding to the Galactic Center scale. This is consistent with previous studies showing that current-generation detectors are sensitive to CCSNe in the Galactic Center~\citep{gossan16observing, Szczepanczyk21Detecting}, making such inference scenarios realistic. Next-generation detectors substantially extend the reach of reliable parameter estimation. With ET, accurate recovery of physical parameters is possible out to distances of $\sim 200$\,kpc, while CE further extends this range to $\sim 350$\,kpc. 

To further quantify the distance-dependent performance across realistic astrophysical targets, Table~\ref{tab:galaxy_scores} summarizes the $R^2$ scores for 34 nearby galaxies and dwarf galaxies spanning distances from 20\,kpc to 520\,kpc. We find that the performance of the current A+ detector degrades rapidly with increasing distance. Adopting a reliability threshold of $R^2 \geq 0.8$, $T/|W|$ remains reliably recoverable only out to the distance of Boötes (66\,kpc), while $D\Delta h$ can be estimated up to Coma Berenices (42\,kpc). The estimation of $f_{\mathrm{peak}}$ is the most challenging, yielding $R^2 < 0.8$ even at the nearest distances considered. In contrast, the third-generation detectors maintain substantially higher predictive performance. For ET, $T/|W|$ remains reliably recoverable throughout the entire distance range considered, while $f_{\mathrm{peak}}$ and $D\Delta h$ remain above the reliability threshold out to Canes Venatici~I (220\,kpc) and Leo~V (180\,kpc), respectively. CE provides the strongest overall performance, with all three parameters remaining accurately recoverable ($R^2 > 0.8$) out to the most distant galaxy in the sample, Pisces~V (520\,kpc).

Overall, these results demonstrate that the dependence of regression performance on distance is strongly dictated by detector sensitivity. The transition from 2G to 3G observatories leads to a dramatic expansion of the parameter estimation horizon, significantly improving the prospects for extragalactic CCSN inference.

\begin{table*}[ht!]
\centering
\caption{Predictive performance ($R^2$ score) for the estimation of $T/|W|$, $f_\mathrm{peak}$, and $D\Delta h$ across 34 nearby galaxy distances. Results are shown for three detector configurations: A+, ET and CE. Bold values indicate the reliability horizon (last distance where $R^2 \geq 0.8$); where $R^2$ never reaches $0.8$, the nearest distance is shown in bold instead.}
\label{tab:galaxy_scores}
\resizebox{0.9\textwidth}{!}{
\begin{tabular}{l | c | ccc | ccc | ccc}
\toprule
& & \multicolumn{3}{c}{A+} & \multicolumn{3}{c}{ET} & \multicolumn{3}{c}{CE} \\
\cmidrule(lr){3-5} \cmidrule(lr){6-8} \cmidrule(lr){9-11}
Galaxy name & Distance (kpc) & $T/|W|$ & $f_\mathrm{peak}$ & $D\Delta h$ & $T/|W|$ & $f_\mathrm{peak}$ & $D\Delta h$ & $T/|W|$ & $f_\mathrm{peak}$ & $D\Delta h$ \\
\midrule
Sagittarius & 20.0 & 0.97 & {\bf 0.73} & 0.94 & 0.99 & 0.90 & 0.99 & 0.99 & 0.91 & 0.99 \\
Segue 1 & 23.0 & 0.96 & 0.75 & 0.92 & 0.99 & 0.93 & 0.99 & 0.99 & 0.93 & 0.99 \\
Tucana III & 25.0 & 0.96 & 0.69 & 0.91 & 0.99 & 0.93 & 0.99 & 0.99 & 0.93 & 0.99 \\
Hydrus I & 27.6 & 0.95 & 0.68 & 0.89 & 0.99 & 0.91 & 0.99 & 0.99 & 0.91 & 0.99 \\
Carina III & 27.8 & 0.95 & 0.71 & 0.90 & 0.99 & 0.93 & 0.99 & 0.99 & 0.93 & 0.99 \\
Triangulum II & 30.0 & 0.95 & 0.67 & 0.88 & 0.99 & 0.92 & 0.99 & 0.99 & 0.93 & 0.99 \\
Reticulum II & 32.0 & 0.94 & 0.66 & 0.88 & 0.99 & 0.92 & 0.99 & 0.99 & 0.92 & 0.99 \\
Ursa Major II & 34.7 & 0.94 & 0.65 & 0.85 & 0.99 & 0.92 & 0.99 & 0.99 & 0.92 & 0.99 \\
Segue 2 & 35.0 & 0.94 & 0.63 & 0.86 & 0.98 & 0.91 & 0.99 & 0.99 & 0.91 & 0.99 \\
Carina II & 36.2 & 0.94 & 0.62 & 0.86 & 0.99 & 0.91 & 0.99 & 0.99 & 0.91 & 0.99 \\
Coma Berenices & 42.0 & 0.92 & 0.59 & {\bf 0.82} & 0.99 & 0.92 & 0.99 & 0.99 & 0.92 & 0.99 \\
Willman I & 45.0 & 0.91 & 0.56 & 0.78 & 0.99 & 0.90 & 0.99 & 0.99 & 0.91 & 0.99 \\
Large Magellanic Cloud & 50.0 & 0.89 & 0.54 & 0.76 & 0.99 & 0.91 & 0.99 & 0.99 & 0.92 & 0.99 \\
Tucana II & 58.0 & 0.87 & 0.52 & 0.71 & 0.99 & 0.90 & 0.99 & 0.99 & 0.91 & 0.99 \\
Small Magellanic Cloud & 60.0 & 0.86 & 0.50 & 0.69 & 0.99 & 0.91 & 0.99 & 0.99 & 0.92 & 0.99 \\
Boötes & 66.0 & {\bf 0.83} & 0.50 & 0.64 & 0.99 & 0.90 & 0.99 & 0.99 & 0.91 & 0.99 \\
Draco & 80.0 & 0.78 & 0.37 & 0.55 & 0.98 & 0.87 & 0.99 & 0.98 & 0.88 & 0.99 \\
Sculptor & 84.3 & 0.72 & 0.41 & 0.52 & 0.98 & 0.90 & 0.99 & 0.99 & 0.91 & 0.99 \\
Horologium I & 87.0 & 0.73 & 0.38 & 0.48 & 0.98 & 0.90 & 0.99 & 0.99 & 0.91 & 0.99 \\
Sextans & 90.0 & 0.71 & 0.35 & 0.46 & 0.98 & 0.87 & 0.98 & 0.98 & 0.89 & 0.99 \\
Ursa Major I & 97.2 & 0.69 & 0.34 & 0.46 & 0.98 & 0.87 & 0.98 & 0.99 & 0.89 & 0.99 \\
Carina & 100.0 & 0.67 & 0.26 & 0.40 & 0.98 & 0.89 & 0.98 & 0.99 & 0.90 & 0.99 \\
Aquarius II & 107.9 & 0.61 & 0.27 & 0.36 & 0.98 & 0.87 & 0.98 & 0.99 & 0.89 & 0.99 \\
Grus I & 120.0 & 0.52 & 0.18 & 0.27 & 0.98 & 0.88 & 0.98 & 0.99 & 0.91 & 0.99 \\
Fornax & 140.0 & 0.43 & 0.11 & 0.22 & 0.98 & 0.85 & 0.98 & 0.98 & 0.89 & 0.98 \\
Hercules & 150.0 & 0.35 & 0.07 & 0.12 & 0.98 & 0.84 & 0.97 & 0.98 & 0.88 & 0.98 \\
Leo IV & 160.0 & 0.30 & 0.05 & 0.10 & 0.98 & 0.84 & 0.97 & 0.98 & 0.87 & 0.98 \\
Leo V & 180.0 & 0.22 & -0.01 & 0.07 & 0.98 & 0.82 & 0.96 & 0.98 & 0.88 & 0.98 \\
Pegasus III & 210.0 & 0.09 & -0.10 & -0.05 & 0.97 & 0.79 & 0.96 & 0.98 & 0.85 & 0.98 \\
Canes Venatici I & 220.0 & 0.07 & -0.11 & -0.05 & 0.97 & {\bf 0.82} & 0.95 & 0.98 & 0.88 & 0.97 \\
Leo II & 250.0 & 0.00 & -0.12 & -0.09 & 0.97 & 0.75 & 0.95 & 0.98 & 0.83 & 0.97 \\
Leo I & 260.0 & -0.03 & -0.13 & -0.10 & 0.97 & 0.78 & 0.94 & 0.98 & {\bf 0.85} & 0.97 \\
Phoenix V & 420.0 & -0.21 & -0.21 & -0.19 & 0.94 & 0.67 & 0.89 & 0.97 & 0.75 & 0.95 \\
Pisces V & 520.0 & -0.16 & -0.17 & -0.18 & {\bf 0.93} & 0.63 & {\bf 0.84} & {\bf 0.96} & 0.72 & {\bf 0.93} \\
\bottomrule
\end{tabular}}
\end{table*}

\section{Conclusion}
\label{sec:conclusion}

In this work, we have investigated the prospects for inferring CCSN parameters from the emitted GW signal. We have focused on the bounce and early post-bounce ring-down signals in rotating CCSNe. Using classical machine learning models, we have performed regression-based inference of key physical parameters, including $T/|W|$, $f_{\mathrm{peak}}$, and $D\Delta h$, under increasingly realistic observational conditions. In particular, we have assessed the impact of detector sensitivity, source distance, bounce time uncertainty, equation of state, progenitor variability, and source orientation on parameter recovery.

Among the various models considered, we have found that gradient boosting methods show the best overall performance, with XGB emerging as the most accurate across all tested configurations. Linear models such as Ridge and Lasso regression have been found to be insufficient to capture the nonlinear structure of the problem, while SVR and RFR provide the weakest performance overall (cf. Section~\ref{sec:ML}).

Our results demonstrate that bounce time uncertainty is a critical factor for analysis with time-domain representations, leading to a significant degradation in performance when waveforms are not precisely aligned. In contrast, frequency-domain features remain robust under timing uncertainties up to 20\,ms, highlighting their suitability for realistic inference scenarios where the exact bounce time is unknown (cf. Section~\ref{sec:bouncetime}).
We have also found that intrinsic source properties such as EOS type and progenitor mass do not significantly affect parameter estimation performance (cf. Section~\ref{sec:eos}-\ref{sec:mass}). However, source orientation has a non-negligible impact: parameter recovery is most strongly degraded for low inclination angles between the rotation axis and the line of sight ($\theta \lesssim 20^\circ$), where projection effects suppress the observable strain amplitude and reduce the available information content (cf. Section~\ref{sec:inclination}).

In terms of observational reach, our results show that reliable parameter estimation with second-generation detectors such as A+ is limited to Galactic-scale distances, with robust performance up to approximately $\sim 15$\,kpc. Beyond this range, performance rapidly degrades due to decreasing signal-to-noise ratio. Next-generation detectors significantly extend this horizon, with the ET enabling reliable inference out to $\sim 200$\,kpc and the CE further extending it to $\sim 350$\,kpc, demonstrating the potential of third-generation observatories for extragalactic CCSN studies (cf. Section~\ref{sec:distance}).

Despite these encouraging results, several limitations remain. Our analysis is restricted to rotating core-collapse models and short-time windows around bounce, which does not capture full complexity of supernova signals. We also assume single-detector observations with optimal orientation, without considering the benefits of detector networks, which can improve parameter estimation accuracy~\citep{Bruel23Inference}. Addressing these limitations will be the focus of future work.

\begin{acknowledgments}
This research was funded by the Science Committee of the Ministry of Science and Higher Education of the Republic of Kazakhstan (Grant No. AP26103591). EA is partially supported by the Nazarbayev University Faculty Development Competitive Research Grant Program (no. 040225FD4713). 
SN is supported by FCT - Fundação para a Ciência e Tecnologia, I.P. through doctoral scholarship 2025.02005.BD.
SN acknowledges financial support by FCT in the framework of the Strategic Funding UID/04650/2025.
JAF is supported by the Spanish Agencia Estatal de Investigación (grant PID2024-159689NB-C21) funded by MICIU/AEI/10.13039/501100011033 and by FEDER / EU, and by the Generalitat Valenciana (Prometeo Excellence Programme grant CIPROM/2022/49).

AI-based language tools were used to improve the clarity and grammar of the manuscript.

This research has made use of data or software obtained from the Gravitational Wave Open Science Center (gwosc.org), a service of the LIGO Scientific Collaboration, the Virgo Collaboration, and KAGRA.

\end{acknowledgments}

\bibliography{gw_sn}

\end{document}